\documentclass[a4paper,12pt,color,epsfig]{article}
\usepackage{graphicx}
\usepackage{psi-scirep}
\usepackage{fancyhydr}
 \renewcommand{\headrulewidth}{0pt}

\begin{document}
\rm\small
\title
{The first result of the neutrino magnetic moment measurement in
the {\sl GEMMA} experiment.}

\collaboration{}{{\sl $\dagger$ ITEP} (State Science Center,
Institute for
Theoretical and Experimental Physics, Moscow, Russia)\\
{\sl $\ddagger$ JINR} (Joint Institute for Nuclear Research,
Dubna, Russia)}

\author{ A.G.~Beda$^\dagger$, V.B.~Brudanin$^{\ddagger\star}$, E.V.~Demidova$^\dagger$,
V.G.~Egorov$^\ddagger$, M.G.~Gavrilov$^\dagger$,
M.V.~Shirchenko$^\ddagger$, A.S.~Starostin$^{\dagger\star\star}$,
Ts.~Vylov$^\ddagger$}

\abstract{$\star$)\hspace{2mm}  e-mail: Brudanin@nusun.jinr.ru\\
          $\star\star$)  e-mail: Starostin@itep.ru\\[2mm]
 The first result of the neutrino magnetic moment
measurement at the Kalininskaya Nuclear Power Plant ({\sl KNPP})
with the {\sl GEMMA} spectrometer is presented. An
antineutrino-electron scattering is investigated. A high-purity
germanium detector of 1.5 kg placed 13.9 m away from the 3~GW
reactor core is used in the spectrometer. The antineutrino flux is
2.73$\times$10$^{\rm 13}\; \bar\nu_e$/cm$^2$/s. The differential
method is used to extract the $\nu${\em -e} electromagnetic
scattering events. The scattered electron spectra taken in 6200
and 2064 hours for the reactor {\em on} and {\em off} periods are
compared. The upper limit for the neutrino magnetic moment $\mu
_{\nu } < $5.8$\times$10$^{\rm -11}\mu_B$ at 90{\%} CL is derived
from the data processing. }

%$^{a }

%\begin{center}
%$^{b }$
%\end{center}
%\twocolumn
\begin{contribution}

\section{Introduction}

In recent years a series of remarkable indications of the
atmospheric, solar and reactor neutrino oscillations have been
observed. Analysis of the entire experimental data allows a
conclusion that the neutrino has a finite mass and the neutrino
state mixing process to be defined. However, some fundamental
neutrino properties are not determined as yet. One of them is the
neutrino magnetic moment (NMM). For the massive neutrino the
Minimally Extended Standard Model predicts a very small NMM value
which cannot be experimentally observed at present:\vspace*{-2mm}
\begin{equation}\label{eq.SM}
 \mu _\nu = \frac{3\;e\;G_F }{8\;\pi ^2\sqrt 2 } \cdot
m_\nu \approx 3 \cdot 10^{ - 19}\mu _B \cdot \frac{m_\nu }{1\rm
eV}
\end{equation}
Here, $\mu_B$ is the Bohr magneton ($\mu_B = eh/2m_e$) and $m_\nu$
is the neutrino mass. On the other hand, there is a number of
extensions of the theory beyond the Minimal Standard Model where
the NMM could be at a level of $\left({\rm 10}^{-12}\div{\rm
10}^{-10}\right)$~$\mu _B$ irrespective of the neutrino
mass\cite{Voloshin86,Fukugita87,Pakvasa03}. A rough upper limit of
the NMM value can be obtained from the astrophysical
considerations, namely from the parameters of some stars at the
last stage of their evolution when energy is released mainly as a
neutrino flux. In this case, estimation of the mass of the helium
star core at the moment of the outburst, luminosity of White
Dwarfs and the neutrino energy spectrum at the Supernova explosion
provide the astrophysical NMM upper limits in the range of
$\left({\rm 10}^{-12}\div{\rm 10}^{-11}\right)$~$\mu _B$
\cite{Raffelt99-Fukugita94}. Note that these results are model
dependent, and therefore it is rather important to make laboratory
NMM measurements sensitive enough to reach the
$\sim$10$^{-11}\mu_B$ region. It would allow one to test a large
variety of NMM hypotheses beyond the Standard Model.

The NMM measurements were started more than 30 years ago. They use
reactor and solar (anti)neutrinos in laboratory experiments. The
first paper \cite{Reines76} on the observation of {\it
$\bar{\nu}$-e} scattering was published in 1976. The experiment
was carried out by Reines' group at the Savannah River laboratory.
Later, in 1989 P.~Vogel and J.~Engel \cite{Vogel89} derived the
following upper limit for the NMM from these data:
$\mu_\nu<$(2$\div $4)$ \times $10$^{-10}\mu_B$. In 1992 and 1993
the results of the reactor experiments performed at the
Krasnoyarsk reactor by a group from the Kurchatov Institute
\cite{Vidyakin92} and at the Rovno reactor by a group from
Gatchina \cite{Derbin93} were published. Their NMM upper limits
were 2.4$ \times $10$^{ - 10}\mu _{B}$ and 1.9$ \times $10$^{ -
10}\mu _{B}$, respectively.

The recent reactor experiments have been carried out by the {\sl
MUNU} \cite{MuNu05} and {\sl TEXONO} \cite{TEXONO06}
collaborations over the period 2001--2005. The {\sl MUNU} NMM
upper limit is $\mu_\nu < $9.0$\times$10$^{-11}\mu_B$, whereas the
{\sl TEXONO}\hspace{1mm} experiment\hspace{1mm}
produced\hspace{1mm} the currently best\hspace{1mm}
limit:\hspace{1mm} $\mu_\nu$~$<$~7.2$\times$10$^{-11}\mu_B$
(however, the method of the TEXONO data treatment and the result
extraction seems to be questionable). Thus, over a period of
thirty years the sensitivity of reactor experiments increased only
by a factor 3.

The NMM upper limit comparable with the above reactor results was obtained
by the SuperKamiokande col\-la\-bo\-ra\-tion\cite{SK04} in the solar neutrino investigations.
Ana\-lysis of the recoil electron spectrum after the solar neutrino scattering gives
 $\mu_\nu < $1.1$\times$10$^{-10}\mu_B$. It should be mentioned that the NMM results
in the reactor and Sun experiments may be different. Due to the
oscillation process, the flavor composition of the initial
neutrino flux changes during the propagation in vacuum as follows
\cite{Beacom99}:\vspace*{-2mm} {\footnotesize
\begin{equation}\label{eq.oscillations}
|\nu_e(L)\rangle = \sum_k U_{ek}\;\; e^{-iE_\nu L}\;|\nu_k\rangle \;\;,
\end{equation}\\[-3mm]
} where $E_\nu$ is the neutrino energy, $L$ is the distance from
the source, $U_{ek}$ is the unitary mixing matrix element, $k$
labels the mass eigenstates, and the subscript $e$ labels the
initial flavor. Simultaneously with the change of the neutrino
flavor composition, the effective value of the NMM also
changes:\vspace*{-3mm} {\footnotesize
\begin{equation}\label{eq.NMM_oscillations}
\mu_\nu^2(E_\nu, L) = \sum_j \;\left|\sum_k U_{ek}\;\;e^{-iE_\nu L}\;\;\mu_{jk}\right|^{2} ,
\end{equation}\\[-3mm]
} where the summations $j, k$ are over the mass eigenstates, $\mu
_{jk}$ are the magnetic moments or constants that characterize the
coupling of the neutrino mass eigenstates $\nu_j$ and $\nu_k$ to the
electromagnetic field.

Above the flavour mixing in vacuum was considered. But in the case
of solar neutrinos it must be modified for matter-enhanced
oscillation (the resonant MSW effect). Because of this effect the
solar neutrino flux at the Earth surface is a mixture of electron,
muon and (possibly) tau neutrinos. Since the length of
antineutrino propagation in the reactor experiments is rather
short, the neutrino flux includes only electron antineutrinos.
Thus, a comparison of the NMM measurements in the reactor and
solar experiments can be highly pro\-duc\-ti\-ve
\cite{Beacom99,Grimus03-Tortola04}.

In this paper, the first results of the NMM measurement by the
collaboration of the Institute for Theoretical and Experimental
Physics (ITEP, Moscow) and the Joint Institute for Nuclear
Research (JINR, Dubna) are presented. The measurements are carried
out with the {\sl GEMMA} spectro\-meter (\underline{G}ermanium
\underline{E}xperiment on measurement of \underline{M}agnetic
\underline{M}oment of \underline{A}ntineutrino) at the 3~GW
reactor of the Kalininskaya Nuclear Power Plant (KNPP).

\section{Experimental approach}
A laboratory measurement of the NMM is based on its contribution
to the $\nu$-$e$ scattering. For the nonzero NMM the $\nu$-$e$
differential cross section is given \cite{Vogel89} by the sum of
the standard {\sl weak} interaction cross section ($d\sigma_W/dT$)
and the {\sl electromagnetic} one
($d\sigma_{EM}/dT$):\vspace*{-2mm} {\footnotesize
\begin{eqnarray}
\frac{d\sigma_W}{dT} = \frac{G_F^2 m_e }{2\pi}\left[ \left(1 -
\frac{T}{E_\nu}\right)^2\left(1+2\sin^2\theta_W\right)^2 +  \right. \nonumber\\
\left.  + 4\sin^2\theta_W - 2\left(1+2\sin^2\theta_W\right)\sin^2\theta_W
\frac{m_eT}{E_\nu^2}\right]\;\;,
\label{eq.dsW/dT}
\end{eqnarray}
%\vspace*{-2mm}
\begin{equation}
\frac{d\sigma_{EM}}{dT} =\pi r_0^2\cdot \left( {\frac{\mu _\nu }{\mu _B }}
\right)^2\cdot\left( {\frac{1}{T} - \frac{1}{E_\nu }}
\right)\;\;,
\label{eq.dsEM/dT}
\end{equation}
} where $E_\nu$ is the incident neutrino energy, $T$ is the
electron recoil energy, $r_0$ is the electron radius ($\pi
r_0^2=2.495\times10^{-25}$ cm$^2$) and $\theta_W$ is the Weinberg
angle.

In the reactor experiments the pressurized water nuclear reactors
(PWR) of $\sim $3~GW thermal power are used as strong sources of
antineutrinos. In this case one measures the energy spectrum of
electron recoil caused by both weak (W) and electromagnetic (EM)
scattering of the neutrinos, i.e. the sum of these processes, so
that EM plays the role of unremovable background correlated with
the reactor operation.
\begin{figure}[ht]
 \setlength{\unitlength}{1mm}
  \begin{picture}(85,55)(0,0)
   \put(-1,-2){\includegraphics{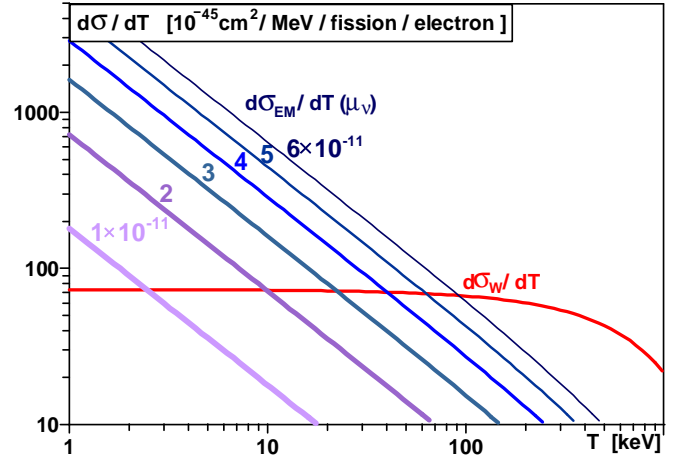}}
  \end{picture}
 \caption{\small Weak (W) and electromagnetic (EM) cross-sections calculated for several
 NMM values.\vspace*{-2mm}}
 \label{Fig.W/EM}
\end{figure}
Figure~\ref{Fig.W/EM} shows the differential cross sections
(\ref{eq.dsW/dT}) and (\ref{eq.dsEM/dT}) averaged over the typical
antineutrino reactor spectrum vs the electron recoil energy. One
can see two important features in this figure. First, because of
an extremely
low cross section, %as in all neutrino experiments,
the problem of the signal-to-background ratio arises, which in its
turn requires strong suppression of all background components.
Second, at low recoil energy ($T\!\ll\!E_\nu$)\hspace{1mm} the
value\hspace{1mm} of $d\sigma_W/dT$\hspace{1mm} becomes constant,
while $d\sigma_{EM}/dT$ increases as $T^{-1}$, so that lowering of
the detector threshold leads to the considerable increase of the
NMM effect with respect to the weak contribution.

More than fifteen years ago M.B.Voloshin and A.S. Sta\-ro\-stin
(ITEP) proposed to search for NMM by means of a low-background
germanium spectrometer (LBGS) similar to ones used in the $\beta
\beta $-decay experiments \cite{Vasenko89} with $^{76}$Ge. The
idea was realized within the project GEMMA by the ITEP-JINR
collaboration in 1997 when the single-crystal Ge(Li) LBGS was
constructed and tested \cite{Beda98}. The LBGS has a low level of
intrinsic noise and additional background suppression in the
energy range below 100~keV due to absorption of the low energy
component of the external radiation by dead layers of the
germanium detector (GD). A combination of these advantages allows
measurements in the energy range $\sim (2 - 100)$~keV. In
addition, the LBGS has a number of characteristics which open the
way to reaching an ultralow level of radiation background
\cite{Aalseth99}. Radioactive impurities in the germanium crystal
do not exceed 10$^{-14}$~g/g, that is 5--7 orders of magnitude
lower than in pure metals. Due to a small size of the GD it can be
easily shielded with such rare and expensive materials as
oxygen-free electrolytic copper, archaeological lead free of
radioactive $^{210}$Pb, etc. One can use NaI crystals as an active
shielding against the charged component of cosmic radiation and
for suppression of the Compton component of external radiative
background.

\section{The GEMMA Spectrometer}

The basic challenge of the experiment is to decrease the
background level of the surface (not {\sl underground}!) setup,
placed close to the reactor, down to the value of (1--2)
events/keV/kg/day. During development of the spectrometer many
different approaches were analyzed in order to solve this problem
\cite{Beda98}. Spectrometers with passive shielding and those with
active shielding were considered. The latter approach was adopted
since it provides better suppression of all background components
under strong cosmic radiation and operation of the nuclear
reactor. The detailed description of the GEMMA spectrometer tested
under the ITEP laboratory conditions is given in
\cite{Beda98,Beda04}. Two important changes were made later during
the assembling of the spectrometer under the KNPP reactor.
Firstly, the four-crystal Ge(Li) detector was replaced by a
one-crystal high-purity germanium detector (HPGe) of 1.5 kg in
mass. Secondly, we abandoned neutron shielding in the form of two
8-cm-thick layers of borated polyethylene, since neutron
background under the reactor was found to be substantially lower
than in the surface laboratory. Besides, removing 16 cm of
polyethylene, we made the spectrometer more compact and thus could
increase the lead shielding against external $\gamma$-radiation.

We use standard CAMAC and NIM modules in the electronic data
acquisition system. The analog part operates in the following way.
The HPGe preamplifier signals (Fig.~\ref{Fig.acq}) are distributed
among five spectroscopic amplifiers. Three of them have the same
gain, but different shaping times ($\tau_1$=2$\mu$s,
$\tau_2$=4$\mu$s, $\tau_3$=12$\mu$s) and take data in the 380 keV
energy range. The forth channel is used for the spectrum
monitoring in the range up to 2.8 MeV. External gates for the
corresponding ADCs are produced by a special peak-sensitive
discriminator connected to a logarithmic amplifier
($\tau_0=1\mu$s). It was found that such a trigger system provides
better linearity in the wide energy range (1.5$\div $ 2800) keV.

\begin{figure}[ht]
 \setlength{\unitlength}{1mm}
  \begin{picture}(85,40)(0,0)
   \put(0,-2){\includegraphics{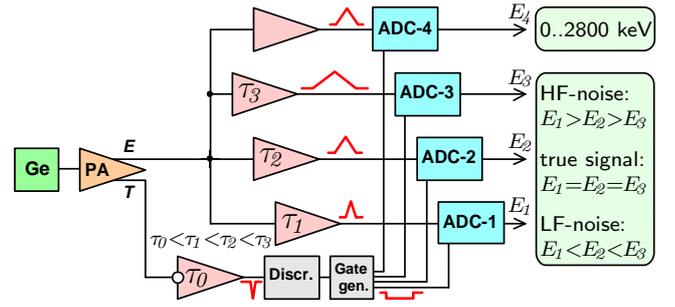}}
    \put(31.5,26.0){\makebox(0,0)[c]{\normalsize $\tau_{\it\! 3}$}}
    \put(34.3,17.5){\makebox(0,0)[c]{\normalsize $\tau_{\it\! 2}$}}
    \put(37.2, 8.7){\makebox(0,0)[c]{\normalsize $\tau_{\it\! 1}$}}
    \put(24.5, 1.5){\makebox(0,0)[c]{\normalsize $\tau_{\it\! 0}$}}
    \put(18.2, 6.5){\makebox(0,0)[l]{\scriptsize\sl
     $\tau_{\it\!0}\!\!<\!\!\tau_{\it\!1}\!\!<\!\!\tau_{\it\!2}\!\!<\!\!\tau_{\it\!3}$}}
    \put(67.0, 36.5){\makebox(0,0)[c]{\scriptsize $E_{\it\! 4}$}}
    \put(67.0, 28.0){\makebox(0,0)[c]{\scriptsize $E_{\it\! 3}$}}
    \put(67.0, 19.5){\makebox(0,0)[c]{\scriptsize $E_{\it\! 2}$}}
    \put(67.0, 11.0){\makebox(0,0)[c]{\scriptsize $E_{\it\! 1}$}}
    \put(69.5, 34.5){\makebox(0,0)[l]{\scriptsize\sf 0..2800 keV}}
    \put(69.5, 13.0){\makebox(0,0)[l]{\parbox{40mm}{\scriptsize\sf
    HF-noise:\\
    {$E_{\it\!1}\!\!>\!\!E_{\it\!2}\!\!>\!\!E_{\it\!3}$\\[2mm]}
    true signal:\\
    {$E_{\it\!1}\!\!=\!\!E_{\it\!2}\!\!=\!\!E_{\it\!3}$\\[2mm]}
    LF-noise:\\
    {$E_{\it\!1}\!\!<\!\!E_{\it\!2}\!\!<\!\!E_{\it\!3}$\\[2mm]}
    }}}
  \end{picture}
 \caption{\small Analog part of the data acquisition system.}
 \label{Fig.acq}
\end{figure}

The use of three amplifiers with different shaping time makes it
possible to suppress low-frequency (LF) and high-frequency (HF)
circuit noises. The first one can be induced by 50-Hz power
supply, as well as microphonics: environmental acoustic noise,
bubbling of liquid nitrogen in the Dewar vessel affecting the
detector cryostat, vibrations in the surrounding equipment, etc.
The sources of the HF circuit noises include pulse radiofrequency
interference due to operation of the PC and electronic units,
fluctuations of the detector dark current, thermal noise of the
preamplifier FET. The LF and HF noise signals are amplified in
different ways for different bands ($\tau_{1}<\tau_{2}<\tau_{3}$).
The relation $E_{1}=E_{2}=E_{3}$ is valid for true signals and
distorted for noise signals, which allows their discrimination
\cite{Garcia92}.

\section{Preparation work}

The GEMMA spectrometer is installed under the second reactor block
of the KNPP. Before the installation, we measured the background
caused by $\gamma$-radiation, thermal neutrons, charged component
of cosmic radiation and radioactive aerosol pollution. The
$\gamma$-radiation background was measured with a portable
germanium detector. An integral level of the radiation background
was an order of magnitude higher than usually indoors, and it was
caused mainly by long-lived fission products: $^{134,137}$Cs and
$^{60}$Co. The measurements were repeated several times during the
reactor ON and reactor OFF periods. The radiation background level
was the same within the statistical errors in both periods. As a
result, it was decided that the existing shielding (15 cm of Pb +
5 cm of Cu + 14 cm of NaI) is quite enough to suppress the
external $\gamma$-background.

The  charged component of cosmic rays (muons) was detected with
two coincident plastic scintillator co\-un\-ters of size
120$\times$120$\times$4 cm$^{3}$. The absolute value and the
angular distribution of the muon flux were measured. Compared to
the surface laboratory conditions, the muon flux was found to be
reduced by a factor of 6 due to passive shielding by the building
structure and the reactor itself. At the same time, the factor of
muon flux suppression was 10 at the angles of $\pm$20$^\circ$ with
respect to the vertical\footnote{In more details, the angular
distribution of cosmic muons was considered, for instance, in
\cite{Bogdanova06}.} and only 3 at the angles of
70$^\circ\div$80$^\circ$, which corresponds to 70 and 20 meters of
water equivalent, resp. The above passive shielding completely
eliminates the hadronic component of the primary cosmic rays.

Neutrons were detected with a set of proportional $^3$He counters
enclosed in a polyethylene moderator. The measurements were
relative: the neutron flux at the spectrometer site was compared
to the flux at the surface laboratory. At the surface one can
normally observe secondary atmospheric neutrons and tertiary
neutrons due to muon capture, the fraction of the tertiary
neutrons being (10$\div$20)\% \cite{Skoro92-Wordel96} or about
160~n/g/year. Fast secondary and tertiary neutrons can be detected
after their thermalization in the moderator. At the spectrometer
site the secondary neutrons must be absent because of the upper
passive shielding. The tertiary neutron contribution must decrease
in proportion to the mu\-on flux (by a factor of 6), but some
additional neutrons can originate from the reactor core. The
measurements proved that the neutron flux at the spectrometer site
is reduced by a factor of 30 compared to the surface. These
results correspond to the above muon measurements and indicate
that the reactor neutrons are not significant.

Much attention was paid to the contamination of the experimental
room with a radioactive dust since the background level could
significantly increase if radioactive dust get settled inside the
shielding during the setup assembling. Samples were taken in
different places of experimental room beneath the reactor and
investigated with the GEMMA spectrometer in the ITEP
low-background laboratory. Radioactive nuclides were identified
and their activity was measured. In addition, different methods of
decontamination were investigated. As a result of these
investigations, it was decided to cover the floor with special
plastic. Once this work was done, test measurements were carried
out, and it became clear that radioactive dust contamination was
significantly reduced.

\section{Reactor antineutrinos}

The setup is located under the 3-GW PWR at a distance of 13.9 m
from the center of the reactor core. During the measurements at
the KNPP the energy threshold of GEMMA spectrometer was 2.5 keV.
At such a low threshold the recoil electron spectrum does not
depend on details of the reactor antineutrino spectrum and is
determined by the {\sl total} reactor antineutrino flux
$\Phi_\nu$: {\footnotesize
\begin{equation}\label{eq.f_nu}
 \Phi_\nu  = N_f N_\nu / 4\pi R^2 \;,
\end{equation}
} where $N_f$ is the number of fissions in the reactor core per
second, $N_\nu$ is the antineutrino yield per fission, $R$ is the
distance from the center of the reactor core. $N_f$ can be
expressed in terms of the reactor thermal power $W$ and an average
energy $E_f$ per fission: {\footnotesize
\begin{equation}\label{Eq.N_f(W)}
N_f = W/E_f\;.
\end{equation}
} The energy $E_f$ depends on the reactor type, fuel composition
and time elapsed since the beginning of the reactor cycle. At the
beginning of the one-year reactor cycle the typical fuel
composition is the following\cite{Kopeikin03}: {\footnotesize
\begin{equation}
 \begin{array}{llcr}
  ^{\sf 235}{\sf U }: & \alpha_{5}({\it t\!=\!0})& = &69\%\\
  ^{\sf 239}{\sf Pu}: & \alpha_{9}({\it t\!=\!0})& = &21\%\\
  ^{\sf 238}{\sf U }: & \alpha_{8}({\it t\!=\!0})& = & 7\%\\
  ^{\sf 241}{\sf Pu}: & \alpha_{1}({\it t\!=\!0})& = & 3\%
 \end{array}
\end{equation}
}
As $^{235}$U burns down, the plutonium fissile isotopes $^{239}$Pu
and $^{241}$Pu are produced in the reactor core. The following fuel
composition is considered as ``standard'':
{\footnotesize
\begin{equation}
 \begin{array}{llcr}
  ^{\sf 235}{\sf U }: & \alpha_{5}& = &58\%\\
  ^{\sf 239}{\sf Pu}: & \alpha_{9}& = &30\%\\
  ^{\sf 238}{\sf U }: & \alpha_{8}& = & 7\%\\
  ^{\sf 241}{\sf Pu}: & \alpha_{1}& = & 5\%
 \end{array}
\end{equation}
}
and the corresponding energies $E_{fk}$ are:
{\footnotesize
\begin{equation}
 \begin{array}{llcr}
  ^{\sf 235}{\sf U }: & E_{f5}& = &201.9 \;\;{\rm MeV\;/\;fission}\\
  ^{\sf 239}{\sf Pu}: & E_{f9}& = &210.0 \;\;{\rm MeV\;/\;fission}\\
  ^{\sf 238}{\sf U }: & E_{f8}& = &205.5 \;\;{\rm MeV\;/\;fission}\\
  ^{\sf 241}{\sf Pu}: & E_{f1}& = &213.6 \;\;{\rm MeV\;/\;fission}
 \end{array}
\end{equation}
}
The average $E_f$ value is
{\footnotesize
\begin{equation}\label{Eq.E_f_aver}
 E_f =\sum_k \alpha_k E_{fk} = 205.3\;\;{\rm
 MeV\;/\;fission}.
\end{equation}
}
From (\ref{Eq.N_f(W)})--(\ref{Eq.E_f_aver}) one can get the
following $N_f$ value for the 3 GW power:
{\footnotesize
\begin{equation}\label{Eq.N_f(3GW)}
 N_f(3\;{\rm GW}) = 9.14\times 10^{19}\;{\rm fissions\;/\;s}\;.
\end{equation}
} The antineutrino yield per fission $N_\nu$ can be represented as
a sum of two components {\footnotesize
\begin{equation}
 N_\nu = N^F + N^C\; ,
\end{equation}
} where $N^F$ corresponds to the antineutrinos produced in
$\beta$-decay of the $^{235}$U, $^{239}$Pu, $^{238}$U and
$^{241}$Pu fission products. These values were calculated several
times, the results being in agreement to $\sim $1\%. Averaging
over three works \cite{Kopeikin80}--\cite{Kopeikin97}, we get
{\footnotesize
\begin{equation}
 \begin{array}{llcr}
  ^{\sf 235}{\sf U }: & N^F_{235}& = &6.12 \;\;\bar\nu_e\;/\;{\rm fission}\\
  ^{\sf 239}{\sf Pu}: & N^F_{239}& = &5.53 \;\;\bar\nu_e\;/\;{\rm fission}\\
  ^{\sf 238}{\sf U }: & N^F_{238}& = &7.11 \;\;\bar\nu_e\;/\;{\rm fission}\\
  ^{\sf 241}{\sf Pu}: & N^F_{241}& = &6.36 \;\;\bar\nu_e\;/\;{\rm fission}
 \end{array}
\end{equation}
}
and the final value is\vspace*{-2mm}
{\footnotesize
\begin{equation}
N^F=\sum_k \alpha_k N^F_k =6.0 \;\;\bar\nu_e\;/\;{\rm fission}.
\end{equation}
}

The second component $N^C$ corresponds to the antineutrinos emitted
in $\beta $-decay of the nuclides produced due to the neutron
capture by $^{238}$U:\vspace*{-2mm}
{\footnotesize
\begin{equation}
^{\sf 238}{\sf U} (n,\gamma )^{\sf 239}{\sf U}
\;\stackrel{\beta}{\longrightarrow}{}\; ^{\sf 239}{\sf Np}
\;\stackrel{\beta}{\longrightarrow}{}\; ^{\sf 239}{\sf Pu}
\end{equation}
\vspace*{-2mm}
}

The value of $N^C$ equals to 1.2 and was first derived in
\cite{Bakalyarov96-Kopeikin97} with a precision of 5\%. Finally,
for the total number of antineutrinos per fission $N_\nu$ we get
7.2, and substitution of the $N_\nu$ and $N_f$ values into
(\ref{eq.f_nu}) gives the following antineutrino flux at the GEMMA
site: {\footnotesize
\begin{equation}\label{eq.Nu_flux}
\Phi_\nu = 2.73\times 10^{13} \quad {\rm \bar\nu_e/cm^2/s}.
\end{equation}
}

\section{The measurement and preli\-mi\-nary data processing}

In order to get a recoil electron spectrum, we used a difference
method comparing the spectra measured at the reactor operation
(ON) and shut-down (OFF) periods. In this work we consider $\sim
$1 year measurement from 15.08.2005 to 20.09.2006, including 6200
and 2064 ho\-urs of the reactor ON and OFF periods\footnote{The
OFF period includes the time for the planned replacement of the
fuel elements, as well as the unplanned reactor shutdowns caused
by other reasons.}, respectively.

\begin{figure}[hb]
 \setlength{\unitlength}{1mm}
  \begin{picture}(85,28)(0,0)
   \put(0,-4){\includegraphics{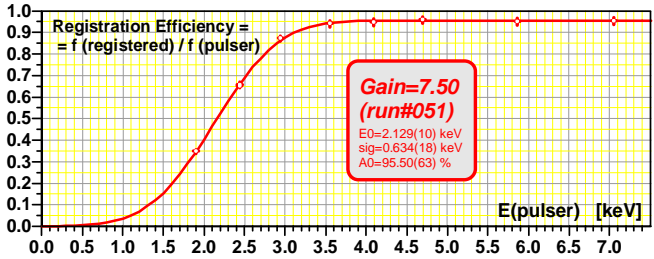}}
  \end{picture}
 \caption{\small Detection efficiency measured with a pulser.}
 \label{Fig.threshold}
\end{figure}

During the measurements, the signals of the HPGe detector,
anti-compton NaI shielding and outer anti-cos\-mic plastic
counters were collected. The energy threshold of the "hard"
trigger ({\sl Discriminator} in Fig.~\ref{Fig.acq}) was as low as
1.5 keV. Detection efficiency just above the threshold was checked
with a pulser (Fig.~\ref{Fig.threshold}). The neutrino flux
monitoring in the ON period was done via the reactor thermal power
measured with the 0.7\% accuracy. The setup was in continuous
operation during all the period mentioned above. Some loss of the
measurement time was caused by occasional failures of the
acquisition electronics, regular liquid nitrogen filling
(increasing the microphonic noise) and sporadic appearance of
additional $\gamma$-background, mainly from $^{133}$Xe (E$\gamma $
= 81~keV) and $^{135}$Xe (E$\gamma $ = 250~keV), due to a failure
of the setup gas tightness. Finally, the data processing was based
on 5184 hours (reactor ON) and 1853 hours (reactor OF).

\begin{figure}[t]
 \setlength{\unitlength}{1mm}
  \begin{picture}(85,64)(0,0)
   \put(-1,-5){\includegraphics{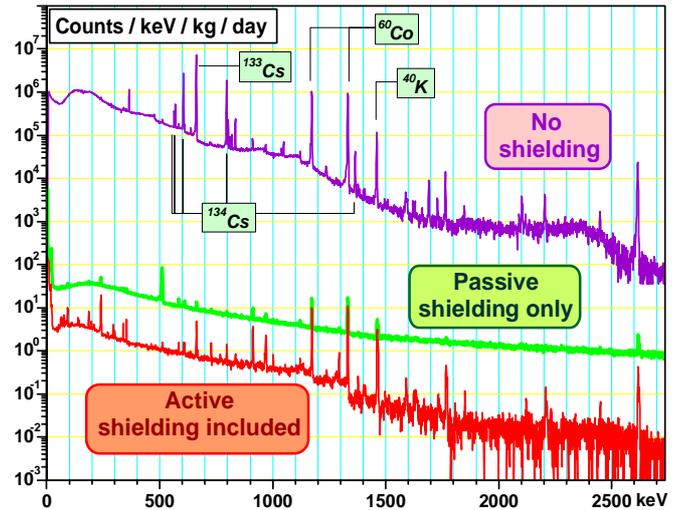}}
  \end{picture}
 \caption{\small Background suppression with shielding.}
 \label{Fig.shld}
\end{figure}

The data processing starts with identification and subsequent
rejection of background events. At the same time, the efficiency
of all the passive and active shielding components was analyzed.
The main effect of the external radiation background suppression
was achieved due to combined passive shielding (PS). The factor of
background suppression in the energy range from 2 keV up to 200
keV exceeded four orders of magnitude. A veto produced by the
inner and outer active shielding signals (NaI and plastic
counters) provides background suppression by one more order of
magnitude (Fig.~\ref{Fig.shld}).

In order to reduce the microphonic and electronic circuit noise,
the simplest Fourier analysis was used. To compare the pulse
heights from three amplifiers (Fig.~\ref{Fig.acq}), three 2D plots
are built: ($E_1$ vs $E_2$), ($E_2$ vs $E_3$) and ($E_3$ vs
$E_1$). An example of such a plot is shown in
Fig.~\ref{Fig.Matrix}.

\begin{figure}[hb]
 \setlength{\unitlength}{1mm}
  \begin{picture}(85,80)(0,0)
   \put(0,-3){\includegraphics{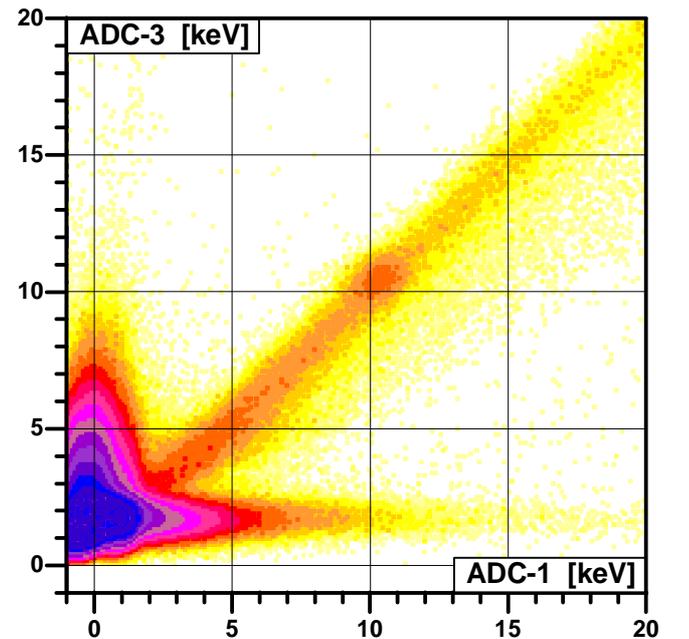}}
  \end{picture}
 \caption{\small Example of the Fourier analysis.}
 \label{Fig.Matrix}
\end{figure}

All true events fall into a diagonal (within the energy
resolution), whereas the events caused by microphonic and
electronic noise are distributed in a different way. The result of
circuit noise suppression with the Fourier analysis is presented
in Fig.~\ref{Fig.Fourier}. It can be seen that the actual energy
threshold can be decreased down to 3 keV.

\begin{figure}[ht]
 \setlength{\unitlength}{1mm}
  \begin{picture}(85,48)(0,0)
   \put(0,-2){\includegraphics{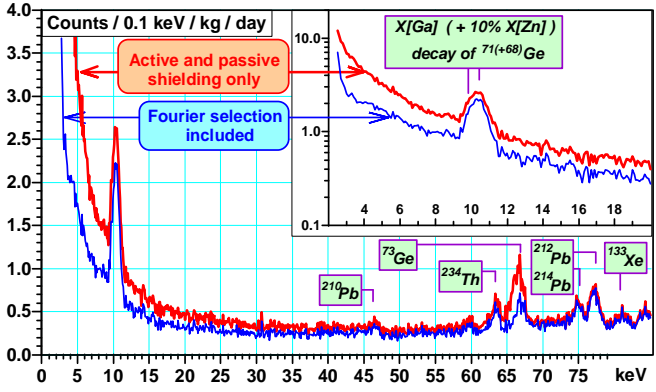}}
  \end{picture}
 \caption{\small Noise suppression with the Fourier analysis.}
 \label{Fig.Fourier}
\end{figure}

After the selection of true signals, ON and OFF energy spectra
normalized to the measurement live time were constructed. The
normalization takes into account the PC dead time (normally less
than 1\%) and random coincidences with the veto signals (about
4\%). The statistics loss caused by the Fourier analysis ($<$1\%)
was estimated from the difference of the background $\gamma $-line
intensities.

Energy spectra shown in Figs.~\ref{Fig.shld} and \ref{Fig.Fourier}
give an idea of the background level and its origin. As is seen
from the figures, the count rate in the energy range from 30 to 45
keV was $\sim$ 2 events/keV/kg/day. Table~1 shows the most intense
$\gamma$-lines measured with the GEMMA spectrometer at the ITEP
laboratory and at the KNPP.

Most of the $\gamma $-lines originate from long lived fission
products ($^{137}$Cs, $^{134}$Cs, $^{60}$Co), natural $^{238}$U
and $^{232}$Th chains, and $^{40}$K. One can see that the U-chain
background ($^{214}$Bi and $^{214}$Pb lines) at the ITEP and KNPP
is the same, whereas the Th-chain and $^{40}$K background is
higher at KNPP. It can be explained by the contamination of the
cryostat components in the process of replacing Ge(Li) detector by
the HPGe detector before the beginning of the measurements. The
presence of $^{60}$Co and $^{137}$Cs $\gamma$-lines in the KNPP
spectrum is caused by the spectrometer contamination during the
setup assembling at the KNPP.

Several $\gamma$-lines originate from the Ge detector activation
by thermal neutrons producing $^{75}$Ge, $^{77m}$Ge and $^{77}$As
$\beta$-active isotopes. A neutron shielding of borated
poly\-ethy\-lene was not used in KNPP measurements. In spite of
this fact, the neutron-induced background there was even lower
than at the ITEP laboratory. The estimated contribution of the
thermal neutrons to the background was about 2{\%}.

\begin{figure}[bh]
 \setlength{\unitlength}{1mm}
  \begin{picture}(85,42)(0,0)
   \put(0,-3){\includegraphics{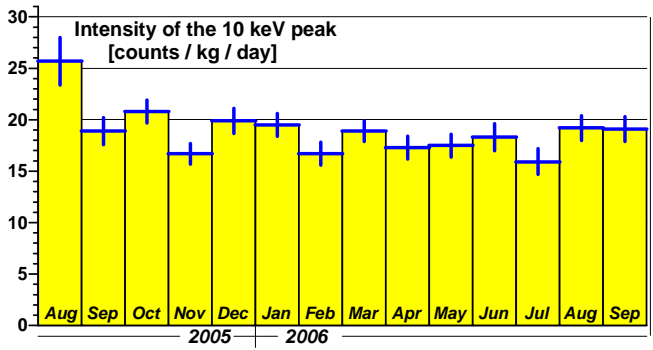}}
  \end{picture}
 \caption{\small Time evolution of the 10 keV peak.}
 \label{Fig.10keV}
\end{figure}

The $\gamma $-line at 10.37 keV can be due to the germanium
detector activation by both thermal and fast cosmogenic neutrons.
In the former case it is due to the $(n,\gamma)$ reaction followed
by $\varepsilon$ capture
\begin{equation} \label{eq.10.37}
\begin{array}{cl}
^{70}{\rm Ge}(n,\gamma)&\!\!\!\!\!^{71}{\rm Ge}\\
& \downarrow  \rule[-3mm]{0mm}{6mm} T_{1/2}=11.4 \;{\rm d}\\
&\!\!\!\!\!^{71}{\rm Ge}\stackrel{\varepsilon}{\longrightarrow}\; ^{71}{\rm Ga}+X({\rm Ga})
\end{array}
\end{equation}
emitting an Auger and/or X-ray cascade with the total energy
deposit of 10.367 keV.

\begin{table}[t]
{\footnotesize
\begin{tabular}{|r|c|r|r|}
%{|p{58pt}|p{160pt}|p{159pt}|}
\hline
Energy& &\multicolumn{2}{|c|}{Intensity}\\
$[$ keV$]$ &Origin of the line&\multicolumn{2}{|c|}{$[$ cnts / kg / day $]$}\\ \cline{3-4}
\rule{0mm}{4mm} &  & $@$ KNPP & $@$ ITEP\\ \hline
\parbox{ 8mm}{\rule{0mm}{4mm}10.37\\
              \rule[-1mm]{0mm}{5mm}10.30}&
\parbox{32mm}{\rule{0mm}{4mm}$^{70}$Ge(n,$\gamma$)$^{71}$Ge, 11.4 d\\
              \rule[-1mm]{0mm}{5mm}$^{70}$Ge(n,3n)$^{68}$Ge, 271 d} &
$\!\!\!\!\!\left.\rule[-3mm]{0mm}{6mm}\right\}$17.5$\pm$0.4&
--- \rule{3mm}{0mm}\\ \hline
\rule{0mm}{4mm}  46.50 & $^{210}$Pb                             &  1.2$\pm$0.1 &--- \rule{3mm}{0mm}\\ \hline
\rule{0mm}{4mm}  63.30 & $^{234}$Pa                             &  3.5$\pm$0.2 &--- \rule{3mm}{0mm}\\ \hline
\rule{0mm}{4mm}  66.70 & $^{72}$Ge(n,$\gamma$)$^{73m}$Ge        &  4.5$\pm$0.2 & 5.5$\pm$1.5 \\ \hline
\rule{0mm}{4mm}  74.80 & $^{214}$Pb                             &  2.8$\pm$0.1 & 5.3$\pm$1.9 \\ \hline
\rule{0mm}{4mm}  77.10 & $^{214}$Pb                             &  5.4$\pm$0.2 & 7.1$\pm$1.9 \\ \hline
\rule{0mm}{4mm}  92.60 & $^{234}$Pa                             & 11.8$\pm$0.3 &--- \rule{3mm}{0mm}\\ \hline
\rule{0mm}{4mm} 139.90 & $^{74}$Ge(n,$\gamma$)$^{75m}$Ge        &  3.4$\pm$0.2 & 5.0$\pm$2.0 \\ \hline
\rule{0mm}{4mm} 186.20 & $^{226}$Ra                             &  9.0$\pm$0.3 & 2.1$\pm$0.5 \\ \hline
\rule{0mm}{4mm} 198.30 & $^{70}$Ge(n,$\gamma$)$^{71m}$Ge        &  2.3$\pm$0.2 & 3.5$\pm$1.2 \\ \hline
\rule{0mm}{4mm} 238.60 & $^{212}$Pb                             & 29.4$\pm$0.5 & 7.5$\pm$1.6 \\ \hline
\rule{0mm}{4mm} 295.20 & $^{214}$Pb                             &  4.4$\pm$0.2 & 5.7$\pm$1.1 \\ \hline
\rule{0mm}{4mm} 338.30 & $^{228}$Ac                             &  4.7$\pm$0.2 &--- \rule{3mm}{0mm}\\ \hline
\rule{0mm}{4mm} 351.90 & $^{214}$Pb                             &  8.0$\pm$0.3 & 7.5$\pm$0.9 \\ \hline
\rule{0mm}{4mm} 511.00 & Annihilation                           &  1.0$\pm$0.1 & 2.5$\pm$1.0 \\ \hline
\rule{0mm}{4mm} 583.20 & $^{208}$Tl                             &  1.7$\pm$0.1 &--- \rule{3mm}{0mm}\\ \hline
\rule{0mm}{4mm} 604.70 & $^{134}$Cs                             &  1.6$\pm$0.1 &--- \rule{3mm}{0mm}\\ \hline
\rule{0mm}{4mm} 609.30 & $^{214}$Bi                             &  2.4$\pm$0.1 & 2.5$\pm$0.9 \\ \hline
\rule{0mm}{4mm} 661.66 & $^{137}$Cs                             &  8.5$\pm$0.3 &--- \rule{3mm}{0mm}\\ \hline
\rule{0mm}{4mm} 727.30 & $^{212}$Bi                             &  2.2$\pm$0.1 &--- \rule{3mm}{0mm} \\ \hline
\rule{0mm}{4mm} 911.10 & $^{228}$Ac                             &  8.4$\pm$0.3 & $\leq$1.0 \rule{2mm}{0mm} \\ \hline
\rule{0mm}{4mm} 969.00 & $^{228}$Ac                             &  4.7$\pm$0.2 &--- \rule{3mm}{0mm} \\ \hline
\rule{0mm}{4mm}1173.20 & $^{60 }$Co                             & 25.7$\pm$0.5 &--- \rule{3mm}{0mm} \\ \hline
\rule{0mm}{4mm}1332.50 & $^{60 }$Co                             & 30.2$\pm$0.5 &--- \rule{3mm}{0mm} \\ \hline
\rule{0mm}{4mm}1460.80 & $^{40 }$K                              & 13.4$\pm$0.1 & 4.1$\pm$0.5 \\ \hline
\rule{0mm}{4mm}1764.50 & $^{214}$Bi                             &  2.1$\pm$0.1 & 1.8$\pm$0.3 \\ \hline
\rule{0mm}{4mm}2614.50 & $^{208}$Tl                             &  1.8$\pm$0.1 & 0.4$\pm$0.1 \\ \hline
\end{tabular}
}
\caption{Gamma-lines}
\end{table}

In the latter case it is caused by the (n,3n) reaction and is
accompanied with a weak 9.659-keV line from the decay of the
granddaughter $^{68}$Ga nucleus:
\begin{equation}
\begin{array}{cll}
^{70}{\rm Ge}(n,3n)&\!\!\!\!\!^{68}{\rm Ge}\\
& \multicolumn{2}{l}{\downarrow \rule[-3mm]{0mm}{6mm} T_{1/2}=271 \;{\rm d}}\\
&\!\!\!\!\!^{68}{\rm Ge}\stackrel{\varepsilon}{\rightarrow}&\!\!\!\!\!^{68}{\rm Ga}+X({\rm Ga})\\
&& \downarrow \rule[-3mm]{0mm}{6mm} T_{1/2}=1.13 \;{\rm h}\\
&&\!\!\!\!\!^{68}{\rm Ga}\stackrel{\varepsilon}{\rightarrow}\; ^{68}{\rm Zn}+X({\rm Zn})
\end{array}
\label{eq.10.30}
\end{equation}
Branching of the second $\varepsilon$ capture is only about 10\%,
so that a slightly broadened peak with a mean energy of 10.30 keV
should be observed.

Since the intensity of this $\gamma $-line (Fig.~\ref{Fig.10keV}) was almost constant
throughout the whole measurement period, it is believed that the first case (\ref{eq.10.37})
dominates over the second one (\ref{eq.10.30}).

\section{The data processing}

At low electron recoil energy the differential cross sections (\ref{eq.dsW/dT}, \ref{eq.dsEM/dT})
take an asymptotic form:
\begin{equation}\label{eq.dW/dT}
\frac{d\sigma_W}{dT}=1.06\times 10^{-44} {\rm cm^2/MeV}
\end{equation}
\begin{equation}\label{eq.dEM/dT}
\frac{d\sigma_{EM}}{dT}=\frac{2.495}{T}\cdot X\times 10^{-45} {\rm cm^2/MeV}
\end{equation}
where the parameter $X$ stands for the NMM squared in terms of
10$^{-10}$ Bohr magnetons:
\begin{equation}\label{eq.X}
X\equiv \left(\frac{\mu_\nu}{10^{-10}\mu_B}\right)^2 \;.
\end{equation}
To convert the cross sections $d\sigma_{W,EM}/dT$ into the recoil
electron spectra $S_{W,EM}(T)$, one has to multiply
(\ref{eq.dW/dT}, \ref{eq.dEM/dT}) by a luminosity $L$ which
depends on the experimental conditions:
\begin{equation}\label{eq.L}
L=N_e\cdot \Phi_\nu\;,
\end{equation}
where $N_e$=4.0$\times$10$^{26}$ is the number of electrons in the
fiducial volume of the germanium detector and the neutrino flux
$\Phi_\nu$ is given by (\ref{eq.Nu_flux}).

At low recoil energy one must take into account the atomic
electron binding effects \cite{Bakalyarov96-Kopeikin97}. The
energy $q$ transferred to an electron in the antineutrino
inelastic scattering (both weak and electromagnetic) on the
$i$-subshell is
\begin{equation}
q = \varepsilon_i + T
\end{equation}
where $\varepsilon_i$ is the binding energy of the $i$-subshell
\cite{Larkins77}, and $T$ is the kinetic energy of the recoil
electron. If the energy transfer is less than the binding energy
($q<\varepsilon_i$), the electron cannot leave the subshell and be
detected via the ionization mode \cite{Kopeikin98-Fayans01}.
Formally, the spectrum must be corrected \cite{Mikaelyan02} by a
function $R$ (Fig.~\ref{Fig.R(q)}):
\begin{equation}\label{eq.R}
R(q) = \frac{1}{Z}\cdot \sum_i n_i \cdot \theta \left( q - \varepsilon_i\right)\;,
\end{equation}
where $Z$ = 32 for germanium, $n_i$ is the number of electrons at
the $i$-subshell, and the $\theta$ factor is
\begin{equation}\label{eq.theta}
 \theta(q-\varepsilon_i)=\left\{
 \begin{array}{ccl}
  1&{\rm if}& q>    \varepsilon_i\\
  0&{\rm if}& q\leq \varepsilon_i
 \end{array}
 \right.
\end{equation}

After the correction, the recoil energy spectra become as follows:
\begin{equation}\label{eq.S_W,EM}
 \begin{array}{lcl}
  S_{EM}(T)& = & \frac{d\sigma_{EM}}{dT}\cdot R \cdot L \\
  \rule{0mm}{6mm}S_{W}(T) & = & \frac{d\sigma_{W}}{dT} \cdot R \cdot L
 \end{array}
\end{equation}

\begin{figure}[ht]
 \setlength{\unitlength}{1mm}
  \begin{picture}(85,33)(0,0)
   \put(1,-2){\includegraphics{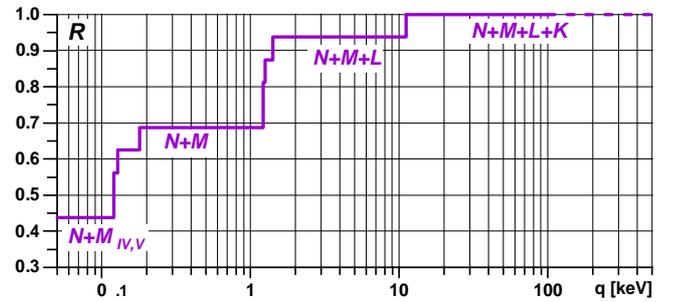}}
  \end{picture}
 \caption{\small Correction for the electron binding energy.}
 \label{Fig.R(q)}
\end{figure}

As is seen in Fig.~\ref{Fig.R(q)}, the elastic and inelastic
spectra are the same if the transferred energy exceeds the
electron binding energy at the K-shell. The correction for bound
electron states in the energy range from 11.1 keV down to the
energy threshold (1.5 keV) comes to only 6.25\%.

The data taken in the reactor ON period include two additional
items ($S_W$ and $S_{EM}$) compared to the data taken in the reactor
OFF period ({\em the other conditions being equal}) :
\begin{equation}
 S_{ON}(T) = S_{OFF}(T) + S_{W}(T) + S_{EM}(T,X)
\end{equation}
In view of (\ref{eq.dW/dT},\ref{eq.dEM/dT}) and (\ref{eq.S_W,EM}),
an experimental estimate of the NMM can be extracted for any
$T$-value:
\begin{equation}\label{eq.X(S)}
  X = \frac{\left( S_{ON} - S_{OFF} - S_{W}\right)\cdot T}
         {2.495\times 10^{-45} \cdot R \cdot L}\;.
\end{equation}

In this work, data were processed in the energy region of interest
(ROI) from 3.0 keV to 61.3 keV with a step of 0.1 keV. To exclude
10.37-keV and 46.5-keV peaks, the above range was divided into
three intervals: (3.0--9.2) keV, (11.2--44.0) keV and (48.0--61.3)
keV (Fig.~\ref{Fig.ON-OFF}). Thus, with the given channel width of
0.1 keV, we get a set of 526 values of $X_i\pm \Delta X_i$, where
$i$=[30..92, 112..440, 480..613] is the channel number and $\Delta
X_i$ is the statistical error in each channel (the upper part of
Fig.~\ref{Fig.ON-OFF}).

\begin{figure}[t]
 \setlength{\unitlength}{1mm}
  \begin{picture}(85,62)(0,0)
   \put(0,-2){\includegraphics{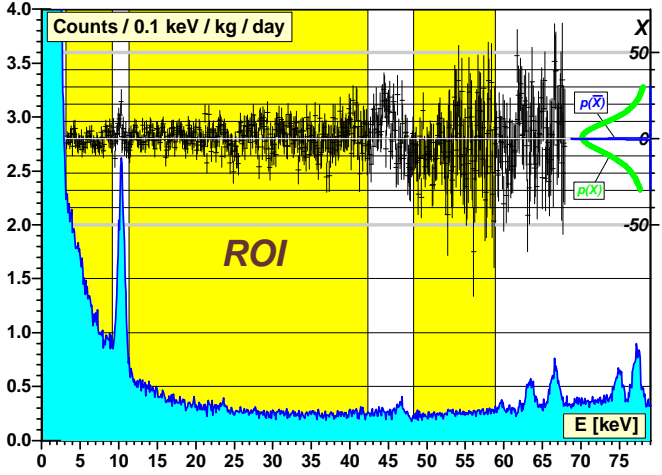}}
  \end{picture}
 \caption{\small Extraction of $X$ from the experimental ROI.}
 \label{Fig.ON-OFF}
\end{figure}

\begin{figure}[b]
 \setlength{\unitlength}{1mm}
  \begin{picture}(85,60)(0,0)
   \put(0,-2){\includegraphics{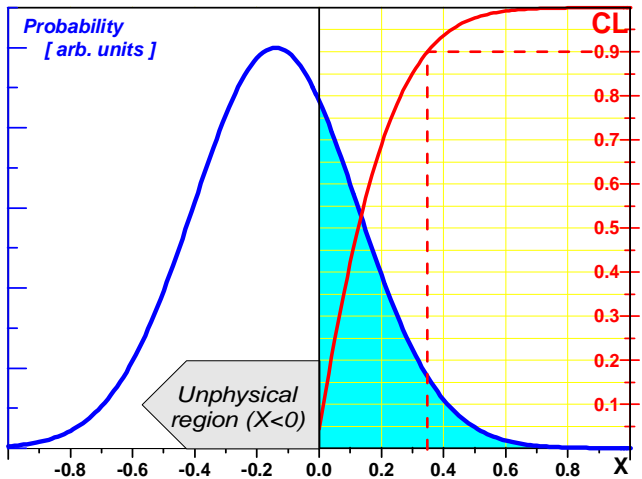}}
  \end{picture}
 \caption{\small Estimation of the $X$ mean value: probability(blue) and confidence level (red).}
 \label{Fig.probability}
\end{figure}

Distribution of $X_i$ is shown in the right part of
Fig.~\ref{Fig.ON-OFF} as $p(X)$. It is a sum of 526 Gaussians with
centers at $X_i$ and variances $\Delta X_i$. The $p(X)$ curve has
a symmetrical Gaussian-like form, which proves the absence of
systematic deviations. A product of these 526 Gaussians represents
the distribution of the probabilistic estimation of the $X$ mean
value ($p\langle X\rangle$); it is shown in
Fig.~\ref{Fig.probability}. The distribution center falls below
zero, in the unphysical region. According to the Bayes strategy
recommended by the Particle Data Group \cite{PDG}, we renormalize
this distribution in such a way that the area under the physical
part of the curve is equal to 1. The integral of this part of the
distribution is shown as a Confidence Level (CL) in
Fig.~\ref{Fig.probability} with a red solid line (dashed red line
is an example of the result extraction at the 90\%CL).

The systematic uncertainty of the result originates from several
sources. The uncertainty of the reactor thermal power for PWR-1000
is about 0.7\%. Another 3\% uncertainty comes from the calculated
value of the antineutrino flux $\Phi_\nu$, which, in its turn,
arises from the uncertainty of the reactor antineutrino spectrum.
Then one must take into account the rejection efficiency and the
efficiency of the Fourier analysis described above; they affect
the final result indirectly, through the intensity of the ON and
OFF spectra, and therefore have been included into the error bars
of Fig.~\ref{Fig.ON-OFF}. Systematic uncertainties and the total
systematic error are presented in Table~2.

\begin{table}[ht]
{\footnotesize
\begin{tabular}{|l|r|r|}\hline
\multicolumn{1}{|c}{Uncertainty source}&
\multicolumn{1}{|c}{\scriptsize $\Delta X/X$}&
\multicolumn{1}{|c|}{\scriptsize $\Delta X$}\\ \hline
Reactor thermal power                                      & 0.7\% & $<0.003$\\
Calculation of $\Phi_\nu$ (ON)                             & 3.0\% & $<0.015$\\
$\Phi_\nu$ during the ON$\leftrightarrow$OFF transitions   & 4.0\% & $<0.020$\\
Correction for the dead+vetoed time                        &       & $<0.010$\\
Correction for the Fourier rejection                       &       & $<0.020$\\
\hline
\multicolumn{2}{|l}{Total systematic error}                        & $<0.035$ \\
\hline
\end{tabular}
}
\caption{\small Systematic uncertainties}
\end{table}

Considering the statistic and systematic errors, the following
limit was derived for $\mu _{\nu }$:
\begin{equation}\label{eq.result}
\mu_\nu<5.8\times10^{-11}\mu_{\rm B}\;\;\;{\rm (90\%CL)}.
\end{equation}
Now the measurement is still in progress, and we expect to improve the GEMMA sensitivity
to the NMM.

\section{Conclusion}

The first result on the neutrino magnetic moment
measurement\hspace{1mm} at the Kalininskaya\hspace{1mm}
Nuclear\hspace{1mm} Power\hspace{1mm} Plant (KNPP) obtained by the
collaboration of ITEP (Mos\-cow) and JINR (Dubna) with the GEMMA
spectrometer is presented. The basis of the spectrometer is a
high-purity germanium detector of 1.5 kg placed at the distance of
$\sim$13.9 m from the center of the 3~GW WPR and surrounded with
NaI active shielding, combined Pb+Cu passive shielding and muon
veto plastic scintillators. The antineutrino flux at the
spectrometer site is $2.73\!\times\!10^{13}\bar\nu_e/{\rm
cm}^2/{\rm s}$. The data were taken during the operation of
reactor (ON period = 6200 hours) and the reactor shutdown (OFF
period = 2064 hours) from 15.08.05 to 20.09.06. When processing
the recoil electron spectra caused by the electromagnetic and weak
interactions, we took into account the effects of the electron
binding in the germanium atoms. The limit on the neutrino magnetic
moment of $\mu_\nu<5.8\times10^{-11}\mu_{\rm B}$ at 90\%CL was
derived from the data analysis.

At present, the data taking is in progress. Simultaneously, we are
preparing the experiment GEMMA II. Within the framework of this
project we plan to use the antineutrino flux of
$\sim5.4\times10^{13} \bar\nu_e /{\rm cm}^2/{\rm s}$, to increase
the mass of germanium detector by a factor of four and to decrease
the background level. These measures will provide the possibility
of achieving the NMM limit at the level of
$1.5\times10^{-11}\mu_{\rm B}$.

\section{Acknowledgments}

The authors thank Directorates of ITEP and JINR for constant
support of this work. The authors are grateful to the
administration of the KNPP and the staff of the KNPP Radiation
Safety Department for permanent assistance in the experiment. We
thank A.V. Salamatin for development of some electronic units and
for participation in the background measurements.

This work is supported by the Russian Federal Agency of Atomic
Energy and by the RFBR project No. 06-02-16281.

\end{contribution}


\begin{thebibliography}{99}
\small
\bibitem{Voloshin86} %[1]
  M.B.~Voloshin, M.I.~Vysotsky and L.B.~Okun,
  JETP {\em (Rus.)} \textbf{64} (1986) 446.
\bibitem{Fukugita87} %[2]
  M.~Fukugita and T.~Yanagida,
  Phys. Rev. Lett. \textbf{58} (1987) 1807.
\bibitem{Pakvasa03} %[3]
  S.~Pakvasa and J.W.F.~Valle, ``Neutrino Properties Before and After
  KamLAND,''  {\bf hep-ph/0301061}.
\bibitem{Raffelt99-Fukugita94} %[4]
  G.G.~Raffelt, Phys. Rep. \textbf{320} (1999) 319;
  M. Fukugita, `` Neutrinos in Cosmology and Astrophysics'',
  preprint Yukawa Institute Kyoto {\bf UITP/K-1086} (1994).
\bibitem{Reines76} %[5]
  F.~Reines, H.S.~Gurr, H.W.~Sobel,
  Phys. Rev. Lett. \textbf{37} (1976) 315.
\bibitem{Vogel89} %[6]
  P.~Vogel and J.~Engel,
  Phys. Rev. {\bf D 39} (1989) 3378.
\bibitem{Vidyakin92} %[7]
  G.S.~Vidyakin {\it et al}.,
  JETP Letters {\em (Rus.)} \textbf{55} (1992) 206.
\bibitem{Derbin93} %[8]
  A.I.~Derbin {\it et al}.,
  JETP Letters {\em (Rus.)} \textbf{57}(1993) 768.
\bibitem{MuNu05} %[9]
  Z.~Darakchieva {\it et al}.,
  Phys. Lett. \textbf{B 615} (2005) 153.
\bibitem{TEXONO06} %[10]
  H.T.~Wong {\it et al}., {\bf hep-ex/0605006}.
\bibitem{SK04} %[11]
 D.W.~Liu {\it et al}. [The Super-Kamiokande Collaboration],
 {\bf hep-ex/0402015}.
\bibitem{Beacom99} %[12]
 J.F.~Beacom and P.~Vogel, {\bf hep-ph/9907383}.
\bibitem{Grimus03-Tortola04} %[13]
 W.~Grimus {\it et al}.,
 Nucl. Phys. {\bf B 648} (2003) 376;
 M.A.~Tortola, {\bf hep-ph/0401135}.
\bibitem{Vasenko89}
 A.A.~Vasenko {\it et al}.,
 Prib. Techn. Exp. {\em (Rus.)} {\bf 2} (1989) 56;
 Mod. Phys. Lett. {\bf A5} (1990) 1299.
\bibitem{Beda98}%[15]
 A.G.~Beda, E.V.~Demidova, A.S.~Starostin, M.B.~Voloshin,
 Yad. Fiz. {\em (Rus.)} {\bf 61} (1998) 72,
 [Phys. At. Nucl. {\bf 61} (1998) 66].
\bibitem{Aalseth99} %[16]
 C.E.~Aalseth {\it et al}.,
 Nucl. Phys. {\bf B} (Proc. Suppl.) {\bf 70} (1999) 236.
\bibitem{Beda04} %[17]
 A.G.~Beda {\it et al}.,
 Yad. Fiz. {\em (Rus.)} {\bf 67} (2004) 1973,
 [Phys. At. Nucl. {\em (Engl. transl.)} {\bf 67} (2004) 1948].
\bibitem{Garcia92} %[18]
 E.~Garcia {\it et al}., Nucl. Phys. {\bf B} ( Proc. Suppl.)
 {\bf 28A} (1992) 286.
\bibitem{Bogdanova06} %[19]
 L.N.~Bogdanova {\it et al}.,
 Yad. Fiz. {\em (Rus.)} \textbf{69} (2006) 1.
\bibitem{Skoro92-Wordel96} %[20]
 G.P.~\v{S}koro {\it et al}.,
 Nucl. Instr. and Meth. {\bf A 316} (1992) 333;
 R. Wordel {\it et al}.,
 Nucl. Instr. and Meth. {\bf A 369} (1996) 557.
\bibitem{Kopeikin03} %[21]
 V.I.~Kopeikin,
 Yad. Fiz. {\em (Rus.)} \textbf{66} (2003) 500
 [Phys. of At. Nucl. {\em (Engl. transl.)} \textbf{66} (2003) 472].
\bibitem{Kopeikin80} %[22]
 V.I.~Kopeikin,
 Yad. Fiz. {\em (Sov.)} \textbf{32} (1980) 62.
\bibitem{Vogel81} %[23]
 P.~Vogel {\it et al}.,
 Phys. Rev. {\bf C 24} (1981) 1543.
\bibitem{Kopeikin97} %[24]
 V.~Kopeikin, L.~Mikaelyan, V.~Sinev,
 preprint ''Kurchatov Institute``
 {\bf IAE-6038/2} (1997).
\bibitem{Bakalyarov96-Kopeikin97} %[25]
 A.M.~Bakalyarov, V.I.~Kopeikin, L.A.~Mikaelyan,
 Yad. Fiz. {\em (Rus.)} \textbf{59} (1996) 1225
 [Phys. of At. Nucl. {\em (Engl. transl.)} \textbf{59} (1996) 1157];
 V.I.~Kopeikin, L.A.~Mikaelyan, V.V.~Sinev,
 Yad. Fiz. {\em (Rus.)} \textbf{60} (1997) 230
 [Phys. of At. Nucl. {\em (Engl. transl.)} \textbf{60} (1997) 172].
\bibitem{Larkins77} %[26]
 F.P.~Larkins,
 At. Data and Nucl. Data Tables \textbf{20} (1977) 311.
\bibitem{Kopeikin98-Fayans01} %[27]
 V.I.~Kopeikin {\it et al}.,
 Yad. Fiz. {\em (Rus.)} \textbf{61} (1998) 2032
 [Phys. of At. Nucl., {\bf 61} (1998) 1859];
 S.~Fayans, L.~Mikaelyan, V.~Sinev,
 Yad. Fiz. {\em (Rus.)} \textbf{64} (2001) 1.
\bibitem{Mikaelyan02} %[28]
 L.A.~Mikaelyan
 Yad. Fiz. {\em (Rus.)} \textbf{65} (2002) 1206.
\bibitem{PDG} %[29]
 Particle Data Group,
 J. Phys. {\bf G 33} (2006) 1.
\end{thebibliography}
\end{document}